\documentclass{article}
\usepackage{graphicx}
\usepackage{amsmath}
\usepackage{fontawesome5}
\usepackage{geometry}
\usepackage{indentfirst}
\usepackage[T1]{fontenc}
\usepackage[utf8]{inputenc}
\usepackage{csquotes}
\usepackage{comment}
\usepackage[table,xcdraw]{xcolor}
\usepackage{listings}
\usepackage{authblk}
\usepackage{lipsum}
\usepackage{minitoc}
\usepackage{booktabs} 
\usepackage{longtable}
\usepackage{array}
\usepackage[justification=justified]{caption}
\usepackage{cite}
\usepackage{lmodern}
\usepackage{orcidlink}
\usepackage{placeins}
\hypersetup{
    colorlinks=true,
    linkcolor=[RGB]{0,51,153},  % Internal links
    urlcolor=[RGB]{0,51,153}    % External URLs
    citecolor=[RGB]{0,51,153},    % Citation links
    filecolor=[RGB]{0,51,153},
    allcolors=[RGB]{0,51,153} 
}

\noptcrule 
\mtcsetfont{parttoc}{section}{\fontsize{7.5pt}{7.5pt}\sffamily}

\newcommand{\github}[1]{%
   \href{#1}{\faGithubSquare}%
}

\title{Beyond Text and Tables: Vision-Language Model Integration in ComProScanner for Extracting Materials Data from Scientific Figures with High Accuracy}

\author[1,2]{Aritra Roy$^*$\orcidlink{0000-0003-0243-9124}}
\author[3]{Enrico Grisan\orcidlink{0000-0002-7365-5652}}
\author[4]{Chiara Gattinoni$^*$\orcidlink{0000-0002-3376-6374}}
\author[1,2]{John Buckeridge$^*$\orcidlink{0000-0002-2537-5082}}

\affil[1]{\,Energy, Materials and Environment Research Centre, London South Bank University, London SE1 0AA, UK.}
\affil[2]{\,School of Engineering and Design, London South Bank University, London SE1 0AA, UK.}
\affil[3]{\,Bioscience and Bioengineering Research Centre, London South Bank University, London SE1 0AA, UK.}
\affil[4]{\,Department of Physics, Kings College London, London WC2R 2LS, UK.}

\date{}

\begin{document}
\maketitle
% \footnotetext{\textsuperscript{*}Corresponding author: blaiszik@uchicago.edu}
% \footnotetext{\textsuperscript{$\dagger$}These authors also contributed substantially to compiling team results and other paper writing tasks}
\begin{center}
\vspace*{-\baselineskip}
\vspace*{-\baselineskip}
$^*$\textbf{\textit{Corresponding authors}}: Aritra Roy (\href{mailto:pgr.aritra.roy@lsbu.ac.uk}{pgr.aritra.roy@lsbu.ac.uk}), Chiara Gattinoni (\href{mailto:chiara.gattinoni@kcl.ac.uk}{chiara.gattinoni@kcl.ac.uk}), John Buckeridge (\href{mailto:j.buckeridge@lsbu.ac.uk}{j.buckeridge@lsbu.ac.uk})
\end{center}
%\thanks{$^\dagger$These authors also contributed substantially to compiling team results and other paper writing tasks}

% ---- Abstract ----
\begin{abstract}
Automated extraction of materials composition-property data from scientific literature has advanced considerably with the development of large language model-based pipelines; however, existing frameworks remain limited to textual and tabular content, overlooking the substantial proportion of quantitative property data reported exclusively in scientific figures. Here, we extend ComProScanner, a fully end-to-end multi-agent framework for automated composition-property database construction, with a native vision-language model (VLM) based figure extraction capability. The extension introduces a \texttt{FigureExtractor} utility for caption-keyword-based figure filtering across all supported publishers, and a \texttt{GraphExtractorTool} agent that passes extracted figures to a configurable VLM to recover composition-property pairs from scientific charts and plots. Four VLMs are selected for evaluation on the basis of the LMArena Diagram leaderboard with an input cost criterion of less than \$1.50 per million tokens. Benchmarking on 50 piezoelectric ceramic articles from the established $d_{33}$ test corpus demonstrates that Gemini-3-Flash-Preview achieves the highest performance with a composition accuracy of 0.97 and a normalised F1 score of 0.97, whilst remaining the most cost-effective model among the four evaluated. We additionally introduce a range-based value error threshold parameter into the evaluation framework, providing a more physically meaningful assessment of numeric property values extracted from figures than exact value matching. These contributions establish VLM-integrated ComProScanner as the first materials-specific, fully automated, multimodal literature mining platform capable of extracting structured composition-property data from text, tables, and figures within a single unified pipeline.
\end{abstract}
 
% ---- 1. Introduction ----
\section{Introduction}\label{sec:intro}
The scale and quality of available materials datasets have become decisive factors in the success of data-driven approaches to materials discovery. Machine learning and deep learning models trained on composition-property data have demonstrated considerable promise for property prediction, materials screening, and inverse design across a broad range of functional material classes~\cite{olivetti2020data,schilling2025text}. Yet the utility of these approaches remains fundamentally constrained by the completeness and representativeness of the underlying databases. Whilst computational repositories such as the Materials Project~\cite{jain2013commentary}, JARVIS-DFT~\cite{choudhary2020joint} and OQMD~\cite{saal2013materials} have made high-throughput DFT data widely accessible, the experimentally measured properties of the vast majority of synthesised materials, including functional ceramics, alloys, polymers and composites, are not captured in any structured database at the same scale. They exist instead as unstructured information embedded across millions of journal articles, accessible only through manual reading. Bridging this gap between the published literature and machine-ready datasets is one of the central challenges facing the materials informatics community.

Automated information extraction from scientific text has a long and productive history in materials science, progressing from rule-based and transformer-based approaches~\cite{swain2016chemdataextractor,mavracic2021chemdataextractor2,huang2022batterybert,kononova2019synthesis,huang2020battery,trewartha2022benchmark} to LLM-based strategies including prompt engineering~\cite{dagdelen2024structured,polak2024chatextract,zimmermann2025reflections2024largelanguage}, fine-tuning~\cite{foppiano2024supermat} and retrieval-augmented generation~\cite{maharana2025rag}. More recently, multi-agent LLM frameworks have been employed to automate various stages of the extraction workflow~\cite{ansari2024eunomia,ghosh2025automated,roy2026knowledgeactionoutcomes2025}; however, these systems still require users to manually download and supply articles, limiting their scalability for large corpora. To address this, we developed ComProScanner~\cite{roy2026comproscanner}, the first fully end-to-end framework that autonomously handles the complete chain from literature search and article retrieval to structured database construction without any human intervention, provided the publishers' Text and Data Mining (TDM) API keys are supplied to the framework. Whilst this approach performs effectively when data are reported in tabular form or stated explicitly in prose, a substantial proportion of materials science literature reports key property values exclusively in graphical form. Although scientific figure extraction has been attempted through specialised digitisation tools~\cite{jiang2022plot2spectra,lee2024matgd,shivasankaran2023lineex,luo2021chartocr} and, more recently, through vision-language model (VLM)-based approaches~\cite{liu2023deplot,liu2023matcha,zheng2024reticular,plotextract2025,alampara2025probing,odobesku2025nanominer}, no automated, publisher-to-dataset framework existed for handling composition-property data from figures within a single end-to-end pipeline.

In this work, we extend ComProScanner with a native VLM-based figure extraction capability by introducing a \texttt{FigureExtractor} utility for caption-keyword-based filtering of relevant figures across all supported publishers, and a \texttt{GraphExtractorTool} CrewAI agent tool that passes extracted figures to a configurable VLM to recover composition-property pairs from scientific charts and plots. Four VLMs are selected for evaluation on the basis of the LMArena Diagram leaderboard~\cite{chiang2024chatbot} with an input cost criterion of less than \$1.5 per million tokens. Performance is assessed on composition-property extraction from figures using a subset of the piezoelectric $d_{33}$ test corpus established in the prior work~\cite{roy2026comproscanner}, focusing exclusively on composition-property extraction as synthesis data was comprehensively evaluated there. We additionally introduce a range-based value error threshold parameter into the evaluation framework, which provides a more physically meaningful assessment of numeric property values read from charts than exact value matching.

% ---- 2. VLM Integration and Model Selection ----
\section{VLM Integration and Model Selection}
\label{sec:vlm_integration}

\begin{figure}[t!]
    \centering
    \includegraphics[width=0.83\linewidth]{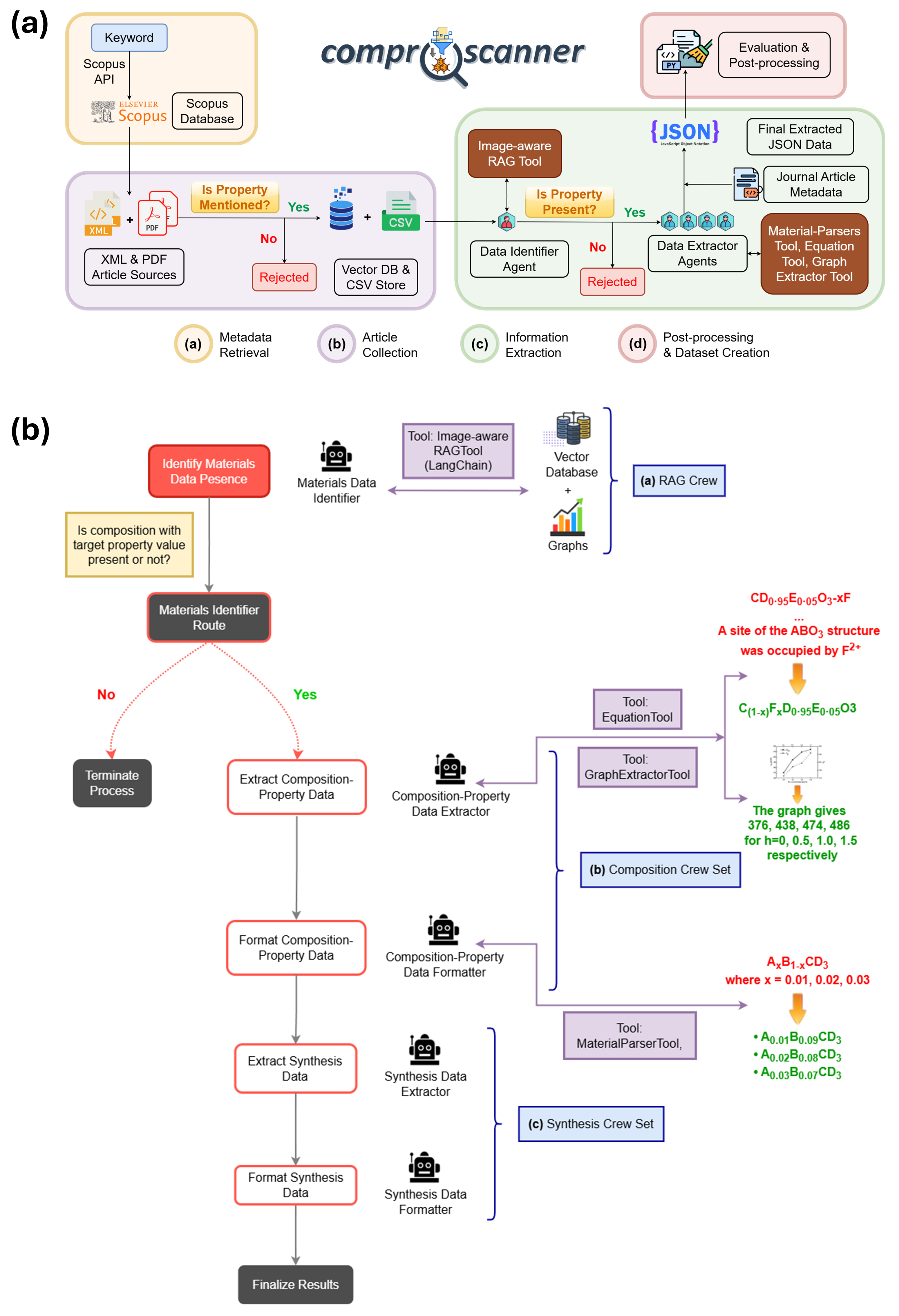}
    \caption{\textbf{(a)} Overall workflow diagram of ComProScanner framework incorporating the GraphExtractorTool and EquationTool. \textbf{(b)} Flow diagram of ComProScanner framework's information extraction process incorporating the image-aware RAGTool, GraphExtractorTool, EquationTool and Material-ParserTool. The detailed descriptions for other components can be found in the original ComProScanner paper~\cite{roy2026comproscanner}.}
    \label{fig:merged_workflows}
\end{figure}

Two complementary mechanisms have been introduced to extend ComProScanner with figure-based extraction. The first is the \texttt{GraphExtractorTool}, a CrewAI \texttt{BaseTool} that, given a DOI, reads all saved figures for that article and passes them to a VLM with a structured extraction prompt, returning composition--property value pairs in the standard ComProScanner JSON schema. The second is an image-aware fallback in \texttt{DataExtractionFlow}: the Materials Data Identifier agent now runs text RAG first; if RAG returns \texttt{no}, the flow checks saved DOI figures via VLM and upgrades the decision to \texttt{yes} when relevant graphical evidence is found. This prevents articles with graph-only data from being silently discarded before extraction begins. A companion \texttt{FigureExtractor} utility handles caption-keyword-based filtering and JPEG conversion, and is shared across all publisher processors. The updated overall workflow is illustrated in Figure~\ref{fig:merged_workflows}(a), and the updated CrewAI-based multi-agent information extraction flow is shown in detail in Figure~\ref{fig:merged_workflows}(b).

\begin{figure}[t!]
    \centering
    \includegraphics[width=1\linewidth]{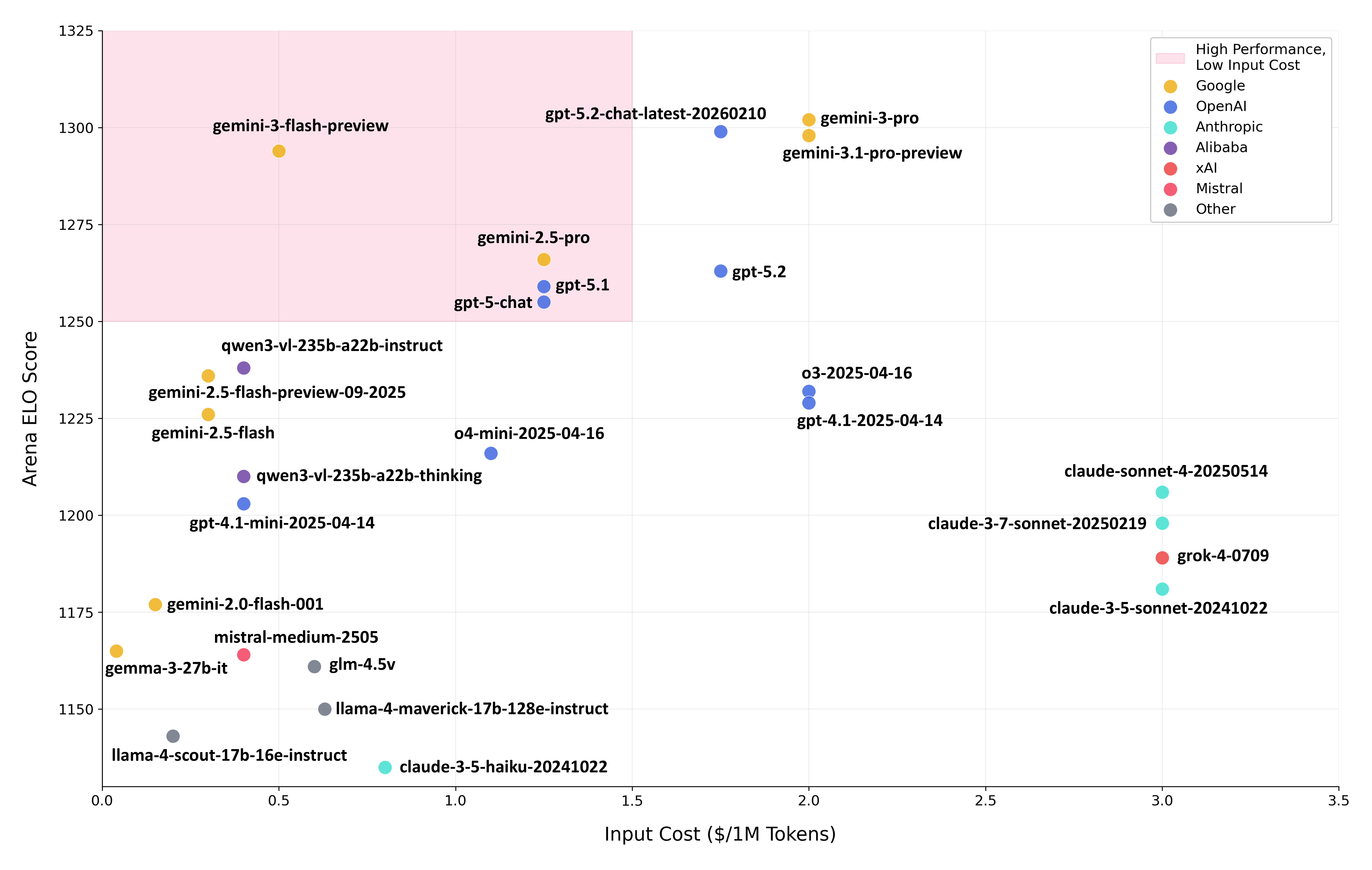}
    \caption{LMArena Leaderboard for VLMs (Diagram category) as of April 2026. The region highlighted in pink indicates the models that were selected for evaluation based on the criteria of having an Arena ELO score of at least 1,250 and an input cost of less than \$1.50 per 1 million tokens.}
    \label{fig:lmarena_vlm_leaderboard}
\end{figure}

Model selection was grounded in the LMArena VLM Leaderboard (Diagram category)~\cite{chiang2024chatbot}, which ranks models by human preference votes on diagram-understanding tasks and reports Arena ELO scores. A critical additional factor was input token cost, as this directly impacts the scalability of the tool for building large datasets. Models were therefore required to satisfy two simultaneous criteria: an Arena ELO score of at least 1,250 and an input cost of less than \$1.50 per million tokens, ensuring a balance between performance and affordability. As of 15 April 2026, this yielded four models for evaluation: Gemini-3-Flash-Preview~\cite{gemini3flash}, Gemini-2.5-Pro~\cite{comanici2025gemini25}, GPT-5-Chat-Latest~\cite{singh2026openaigpt5card} and GPT-5.1~\cite{openai2026gpt51}, as illustrated in Figure~\ref{fig:lmarena_vlm_leaderboard}.

% ---- 3. Results and Discussion ----
\section{Results and Discussion}
\label{sec:results_discussion}
The benchmark was conducted on 50 articles randomly selected from 73 DOIs in the existing piezoelectric ceramic article set that contained related figures. Along with \texttt{GraphExtractorTool}, an \texttt{EquationTool} has been added to the Composition-Property Data Extractor agent (refer to Roy et al.~\cite{roy2026comproscanner} for details) for generating chemical compositions by understanding the element replacement logic and XRD patterns. \texttt{claude-sonnet-4-6} was used for generating the formulae based on the text and XRD patterns. Other settings were kept as the defaults described in the original paper. For saving the figures, a set of keywords related to the piezoelectric coefficient ($d_{33}$) and XRD patterns were used to filter the figures to reduce API costs. However, it should be noted that the filtering process is not perfect and some relevant figures may have been missed; users should be aware of this limitation. The evaluation was performed on the \texttt{composition\_property\_values} field only, using the standard ComProScanner semantic evaluator. Synthesis data including synthesis methods, precursors, and characterisation techniques were excluded from this evaluation as these fields were already evaluated and reported in the prior work~\cite{roy2026comproscanner} and are not affected by graph extraction.

\begin{figure}[t!]
    \centering
    \includegraphics[width=1\linewidth]{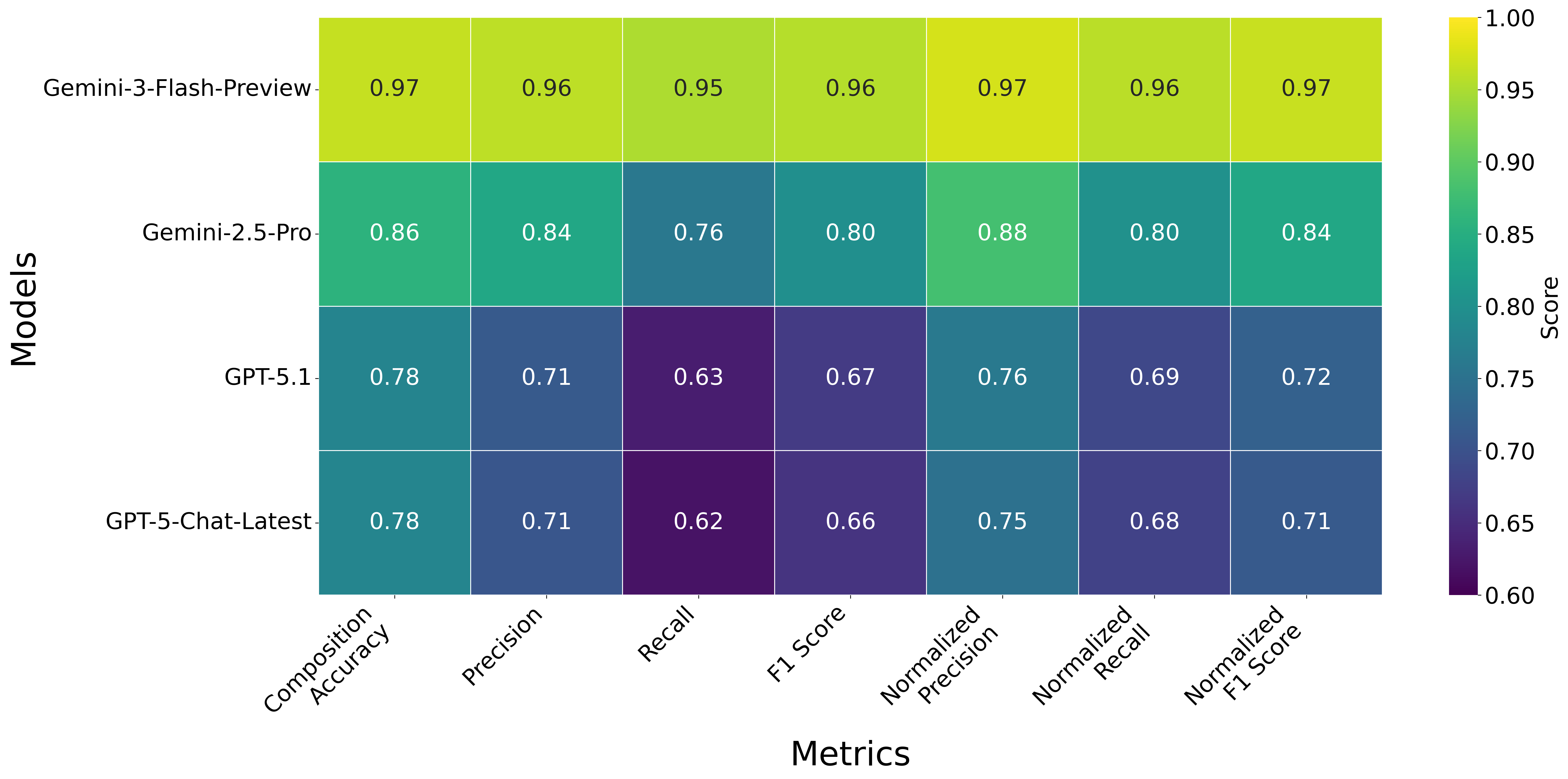}
    \caption{Confusion matrix from semantic evaluation with 1.0 threshold for composition-property data, showcasing all 7 evaluation parameters, such as weight-based composition accuracy, classification metrics (precision, recall and F1-score) and normalised classification metrics (normalised precision, normalised recall and normalised F1-score), across 4 different VLMs used in this study.}
    \label{fig:confusion_matrix}
\end{figure}

Of the 50 selected articles, 48 yielded evaluable composition-property data after extraction and cleaning. One of the remaining two articles, provided by Wiley, was not retrievable even as a PDF. The other article contained environment-dependent $d_{33}$ values which were extracted with `\texttt{--}' and were removed during data cleaning. The evaluation was performed at a strict semantic threshold of 1.0 (exact match) to ensure that the reported metrics reflect precise extraction performance without partial-credit inflation. For $d_{33}$ values, error thresholds of $\pm$0.5, $\pm$1, and $\pm$2 pC/N have been applied for different value ranges. Two complementary classification metric sets are reported, along with weight-based composition accuracy as described in the prior work~\cite{roy2026comproscanner}. The model performance is summarised in the confusion matrix illustrated in Figure~\ref{fig:confusion_matrix}. Gemini-3-Flash-Preview is the strongest performer across all evaluation dimensions, achieving a composition accuracy of 0.97, with absolute precision, recall, and F1 of 0.96, 0.95, and 0.96 respectively, and normalised precision, recall, and F1 of 0.97, 0.96, and 0.97 respectively. This outcome is entirely consistent with its standing on the LMArena Diagram leaderboard, where Gemini-3-Flash-Preview carries a higher Arena ELO score than Gemini-2.5-Pro whilst commanding a substantially lower input cost per million tokens, making it simultaneously the highest-performing and most economical model in this evaluation. Gemini-2.5-Pro performs respectably, with a composition accuracy of 0.86, absolute precision, recall, and F1 of 0.84, 0.76, and 0.80 respectively, and normalised precision, recall, and F1 of 0.88, 0.80, and 0.84 respectively. The notably lower recall relative to precision, a gap of approximately 0.08 in both absolute and normalised settings, suggests the model is more conservative in proposing data points than the Flash variant, consistent with the Pro model's tendency towards cautious reasoning in ambiguous figure layouts. GPT-5-Chat-Latest and GPT-5.1 perform broadly comparably to one another and can be considered together. Both yield a composition accuracy of 0.78, with absolute precision of 0.71, recall of 0.62 and 0.63, and F1 of 0.66 and 0.67 respectively. Normalised metrics follow a similar pattern: precision of 0.75 and 0.76, recall of 0.68 and 0.69, and F1 of 0.71 and 0.72 respectively. Both fall approximately 0.12--0.13 below Gemini-2.5-Pro on normalised F1, indicating difficulty with the full diversity of graphical representations across the corpus. Given this performance gap at similar cost, Gemini-3-Flash-Preview is adopted as the default VLM for the \texttt{GraphExtractorTool}, whilst the \texttt{vlm\_model} parameter remains available for users to override with any LiteLLM-compatible model identifier.

% ---- 4. Additional Improvements ----
\section{Additional Improvements}
\label{sec:additional_improvements}
The following improvements accompany the graph extraction feature. 
\begin{itemize}
    \item A \texttt{value\_error\_thresholds} parameter has been added to both evaluation methods, semantic and agentic, accepting a dictionary mapping \texttt{(min, max)} tuples to absolute error tolerances. The narrowest enclosing range wins, and tuple element order is irrelevant. 
    \item The \texttt{ElsevierArticleProcessor} now accepts a \texttt{SCIENCEDIRECT\_INSTTOKEN} institutional token for off-campus remote access to subscription-based Elsevier articles, forwarded as the \texttt{X-ELS-Insttoken} header.
    \item The \texttt{\_parse\_json\_output()} method now recovers JSON from mixed-text crew outputs via first-brace/ last-brace scanning before falling back to \texttt{ast.literal\_eval()}.
    \item The composition formatter agent now detects and corrects erroneous variable substitution artefacts introduced by the \texttt{MaterialParserTool}.
    \item The \texttt{save\_failed\_pdf\_report} and \texttt{save\_failed\_automated\_report} parameters write tab-separated failure logs for both local PDF and automated publisher workflows respectively.
    \item The \texttt{clean\_data()} function has been improved to handle better cleaning and can now log split-composition resolution statistics and persist filtered and unresolved composition keys to JSON.
    \item The versioning scheme has been switched from SemVer (\texttt{MAJOR.MINOR.PATCH}) to CalVer (\texttt{YYYY.MM.DD}), with the first release under the new scheme being \texttt{2026.05.19}.
\end{itemize}
 
% ---- 5. Conclusions ----
\section{Conclusions}
\label{sec:conclusions}
In this work, we have extended ComProScanner with a native VLM-based figure extraction capability, advancing the framework from a text-and-table mining platform to a fully multimodal, end-to-end materials data extraction pipeline. The introduced \texttt{GraphExtractorTool} and \texttt{FigureExtractor} utility enable automated recovery of composition-property pairs from scientific charts and plots across all supported publishers, addressing a systematic gap in existing literature mining frameworks. Benchmarking across 50 articles from the established piezoelectric $d_{33}$ test corpus demonstrates that Gemini-3-Flash-Preview achieves the strongest performance among the four cost-effective VLMs evaluated, with a composition accuracy of 0.97 and a normalised F1 score of 0.97, whilst simultaneously offering the lowest input cost among the evaluated models. The introduced range-based value error threshold parameter provides a more physically meaningful evaluation of numeric property values extracted from figures than exact value matching, and is applicable to any property domain where graphical data are subject to inherent reading uncertainty. Together, these contributions establish VLM-integrated ComProScanner as the first materials-specific, fully automated, multimodal literature mining platform capable of extracting structured composition-property data from text, tables, and figures within a single unified pipeline, and demonstrate that cost-effective VLMs are sufficiently capable for large-scale deployment in materials informatics workflows.

% ---- Author Contributions ----
\section*{Author Contributions}
AR: conceptualisation, data curation, formal analysis, investigation, methodology, software, validation, writing – original draft. EG: resources, supervision. JB: conceptualisation, formal analysis, funding acquisition, investigation, resources, validation, writing – original draft, writing – review \& editing, supervision. CG: conceptualisation, formal analysis, funding acquisition, investigation, resources, validation, writing – original draft, writing – review \& editing, supervision.
 
% ---- Conflicts of Interest ----
\section*{Conflicts of Interest}
There are no conflicts to declare.
 
% ---- Code and Data Availability ----
\section*{Code and Data Availability}
The benchmark data, model outputs, and evaluation scripts regarding the VLM tests are available in \texttt{examples/vlm\_piezo\_test} folder on the ComProScanner GitHub repository at \url{https://github.com/slimeslab/ComProScanner}.
 
% ---- Acknowledgements ----
\section*{Acknowledgements}
AR and JB thank London South Bank University for financial and
legal support to obtain publisher TDM licences.
CG thanks King's College London for legal support in obtaining
IOP Publishing's TDM licence.
CG was supported by the EPSRC through a New Investigator Award
[grant number UKRI132].

%%%%% Bibliography %%%%%%%%%%%%%%%%%%%%%%%%%%%%%%%%%%%
%\addtocounter{chapter}{1}
% \setcounter{page}{1}
\FloatBarrier
\addcontentsline{toc}{chapter}{References}
\phantomsection
\bibliographystyle{ieeetr}
\bibliography{references}

\end{document}